\documentclass{ifacconf}

\usepackage{amsmath,amssymb,amsfonts}
\usepackage{array}
\usepackage{color}
\usepackage{cuted}
\usepackage{float}
\usepackage{graphicx}
\usepackage{mathrsfs}
\usepackage{multirow}
\usepackage{natbib}
\usepackage{threeparttable}
\usepackage[normalem]{ulem}
\usepackage{url}
\usepackage{verbatim}
\usepackage{textcomp}
\usepackage{xcolor}
\newcommand\dif{\mathrm{d}}

\begin{document}
\begin{frontmatter}

\title{Mechanism Design for Investment Regulation under Herding}

\author[first]{Huisheng Wang}
\author[first]{H. Vicky Zhao}
\address[first]{Department of Automation, Tsinghua University, Beijing, China (e-mail: whs22@mails.tsinghua.edu.cn, vzhao@tsinghua.edu.cn).}

\begin{abstract}
Herding, where investors imitate others' decisions rather than relying on their own analysis, is a prevalent phenomenon in financial markets. Excessive herding distorts rational decisions, amplifies volatility, and can be exploited by manipulators to harm the market. Traditional regulatory tools, such as information disclosure and transaction restrictions, are often imprecise and lack theoretical guarantees for effectiveness. This calls for a quantitative approach to regulating herding. We propose a regulator-leader-follower trilateral game framework based on optimal control theory to study the complex dynamics among them. The leader makes rational decisions, the follower maximizes utility while aligning with the leader's decisions, whereas the regulator designs a mechanism to maximize social welfare and minimize regulatory cost. We derive the follower's decisions and the regulator's mechanisms, theoretically analyze the impact of regulation on decisions, and investigate effective mechanisms to improve social welfare.
\end{abstract}


\end{frontmatter}

\section{Introduction}
Herding is a common phenomenon in financial markets, where investors tend to follow others' decisions rather than relying on their own analysis. This behavior is prevalent on online investment platforms, where influential leaders openly share their decisions and follower investors mimic them (\cite{delellis2017evolution}). While a certain level of herding may be beneficial, e.g., it facilitates information aggregation and accelerates market consensus (\cite{licitra2017single}), excessive herding can lead to severe market distortion, such as asset price bubbles and crashes, thereby amplifying volatility and elevating market risks (\cite{ahmad2022does}). 
These concerns underscore the importance of effective regulatory mechanisms to guide investors' decisions and safeguard market stability.

\subsection{Optimal Investment under Herding}
The optimal investment theory provides a quantitative control-theoretic framework for modeling investors' decisions. The modern portfolio theory assumes that investors make decisions rationally and introduces the static return-volatility optimization problem (\cite{markowitz1952modern}). Based on this, \cite{merton1969lifetime} proposes a dynamic portfolio management model to study the decision during an investment horizon using optimal control theory. Additionally, there have been many other optimal investment problems considering more complex settings, such as consumption and income (\cite{alasmi2024optimal,chen2025optimal}). While these studies provide a foundation for understanding how investors make decisions, they often assume full rationality and neglect the influence of herding. Research on behavioral finance has shown that investors often deviate from rational behavior due to cognitive biases and social influences, among which herding is particularly prominent (\cite{mavruk2022analysis}). The work in \cite{wang2024investment} develops an optimal control-based framework where the follower maximizes his/her utility while tracking the leader's decision. In reality, the regulator aims to maximize social welfare and minimize regulatory costs, whereas individual investors seek to maximize utility. Their objective functions are not necessarily aligned, which calls for effective financial regulatory interventions (\cite{huang2021social}).

\subsection{Financial Regulation and Mechanism Design}
Regulators have long employed tools such as information disclosure and transaction restrictions to mitigate the negative effects of herding and maintain market stability (\cite{armour2016principles}). However, these tools are often imprecise and lack theoretical guarantees for effectiveness (\cite{herring2000optimal}). The mechanism design theory offers a systematic framework for designing rules to achieve desired outcomes, particularly when the regulator lacks access to investors' private information (\cite{borgers2015introduction}). To ensure its effectiveness under such an asymmetric information scenario, the mechanism often satisfies two constraints. The \textit{{individual rationality}} (IR)  constraint guarantees that participation yields no less benefit than opting out (\cite{myerson1979incentive}), while the \textit{{incentive compatibility}} (IC) constraint ensures that investors truthfully report their private information (\cite{roth1977individual}). Mechanism design has been widely applied in domains such as auction formats for truthful bidding (\cite{bei2025optimal}) and call auction rules to improve liquidity (\cite{suen2022call}). It has also played a critical role in financial regulation (\cite{acharya2009restoring}). However, limited work has integrated mechanism design with optimal investment under herding, particularly in regulatory mechanisms to guide investor decisions. 

Regulating herding in financial markets is a complex task. First, traditional regulatory tools often lack theoretical guarantees for effectiveness (\cite{herring2000optimal}). Second, the investor's level of herding is typically private information and unobservable to the regulator, making it difficult to implement targeted interventions, as the appropriate intensity of regulation cannot be easily determined (\cite{aharon2023transparency}). Third, the regulator must balance the regulatory cost against social welfare, defined as the aggregate fund in financial markets, which is both important and challenging to achieve in practice (\cite{goodhart2013financial}). Thus, it is essential to design effective mechanisms that simultaneously address all these issues.

To address these challenges, we formulate a regulator-leader-follower trilateral game problem that integrates optimal investment theory with mechanism design to model and analyze the complex dynamics among investors and regulators. Within this framework, we incorporate an incentive compatibility constraint to ensure that investors truthfully report their levels of herding, and jointly consider the investors' utilities, the regulatory cost, and the social welfare. Furthermore, we study the design of regulatory mechanisms to effectively mitigate herding and theoretically analyze its influence on social welfare.

\section{Problem Formulation}
\label{sec:problem}
In this section, we first introduce the financial market and the leader's decision, and then formulate the follower's and the regulator's optimization problems.

\subsection{The Financial Market}
Following the work in \cite{merton1969lifetime}, to simplify the analysis and to achieve mathematical tractability, we consider a financial market consisting of one risk-free asset whose interest rate is $r\in\mathbb{R}_+$ and one risky asset whose excess return and volatility are $\nu\in\mathbb{R}_+$ and $\sigma\in\mathbb{R}_+$. We assume that there is one regulator, one leader, and one follower in the financial market. In future work, we will study more complex scenarios where $r$, $\nu$, and $\sigma$ are time-varying, and the market involves more than two investors. We assume that the leader makes rational decisions independently to maximize returns and minimize risks, while the follower seeks to maximize returns, minimize risks, and tends to mimic the leader's decisions due to herding. Given that the risky asset offers a positive excess return over the risk-free asset, the regulator's objective function is to encourage the follower to allocate more funds to the risky asset, thus increasing the social welfare (\cite{huang2021social}). 

\subsection{The Leader}
We denote the leader's \textit{{fund}} as $\tilde{\boldsymbol{x}}:=\{\tilde{x}_t\}_{t\in\mathscr{T}}$, where $\mathscr{T}:=[0,T]$ is the finite investment horizon. To simplify the analysis, here, we assume that the leader's initial fund $\tilde{x}_0=0$, and it can be straightforwardly extended to the general cases where $\tilde{x}_0\in\mathbb{R}_+$. We denote the leader's \textit{{decision}} as $\tilde{\boldsymbol{\pi}}:=\{\tilde{\pi}_t\}_{t\in\mathscr{T}}\in\mathscr{P}$, where $\mathscr{P}$ is the set of measurable and absolutely integrable processes. Following the work in \cite{merton1969lifetime}, at each time $t\in\mathscr{T}$, the leader invests $\tilde{\pi}_t$ in the risky asset, and allocates the remaining fund, i.e., $\tilde{x}_t-\tilde{\pi}_t$, to the risk-free asset. 
Thus, his/her fund $\tilde{\boldsymbol{x}}$ can be modeled as
\begin{equation}
    \dif \tilde{x}_t=(r\tilde{x}_t+\nu\tilde{\pi}_t)\dif t+\sigma\tilde{\pi}_t\dif z_t,
\end{equation}
where $\boldsymbol{z}:=\{z_t\}_{t\in\mathscr{T}}$ is a standard Wiener process. 
Following the work in \cite{pratt1978risk}, we assume that the leader's \textit{{utility}}, denoted by $\tilde{\varphi}:\mathbb{R}\to\mathbb{R}$, has the exponential form
\begin{equation}\label{equ:utility}
    \tilde{\varphi}(\tilde{x}_T):=-\tilde{\alpha}^{-1}\mathrm{e}^{-\tilde{\alpha}\tilde{x}_T},
\end{equation}
where $\tilde{\alpha}\in\mathbb{R}_+$ is the leader's \textit{{risk coefficient}}. The utility in \eqref{equ:utility} is increasing, concave, continuous, and differentiable, which facilitates theoretical analysis. A larger $\tilde{\alpha}$ implies that the leader's utility is more sensitive to changes in his/her terminal fund, and thus the leader is more risk-averse. Following the work in \cite{merton1969lifetime}, by maximizing the expected utility of the terminal fund $\mathbb{E}\tilde{\varphi}(\tilde{x}_T)$, we can obtain the leader's optimal decision, which is 
\begin{equation}
    \tilde{\pi}_t:=\frac{\nu}{\tilde{\alpha}\sigma^2}\mathrm{e}^{r(t-T)}.\label{equ:leader}
\end{equation}

\subsection{The Follower}
We denote the follower's fund, decision, utility, and risk coefficient as $\boldsymbol{x}:=\{x_t\}_{t\in\mathscr{T}}$, $\boldsymbol{\pi}:=\{\pi_t\}_{t\in\mathscr{T}}\in\mathscr{P}$, $\varphi:\mathbb{R}\to\mathbb{R}$, and $\alpha\in\mathbb{R}_+$, and we have
\begin{equation}
    \dif x_t=(rx_t+\nu\pi_t)\dif t+\sigma\pi_t\dif z_t,\label{equ:fund_process}
\end{equation}
subject to $x_0=0$. We assume that $\alpha$ is known to the regulator, as investors are typically required to undergo risk assessments for qualifications, and such data is accessible to the regulator (\cite{rajput2013research}). Following the work in \cite{wang2024investment}, we assume that the follower maximizes the expected utility of the terminal fund $\mathbb{E}\varphi(x_T)$ and minimize the deviation between his/her decision $\boldsymbol{\pi}$ and the leader's decision $\tilde{\boldsymbol{\pi}}$ under herding, which is defined as
\begin{equation}\label{equ:delta}
    \delta(\boldsymbol{\pi}):=\frac{1}{2}\int_\mathscr{T}\mathrm{e}^{2r(T-t)}(\pi_t-\tilde{\pi}_t)^2\dif t.
\end{equation}
Therefore, without regulation, the follower's optimal investment problem can be formulated as
\begin{equation}
    \textstyle\sup_{\boldsymbol{\pi}\in\mathscr{P}}\mathbb{E}\varphi(x_T)-\eta \delta(\boldsymbol{\pi}),\label{equ:without_regulation}
\end{equation}
where $\eta\in\mathbb{R}_+$ is the follower's \textit{{herd coefficient}}. A larger $\eta$ implies that the follower has a stronger tendency to align with the leader's decision. The herd coefficient $\eta$ can be reflected in the frequency with which the follower adjusts towards the leader's decision (\cite{bikhchandani1992theory}). For clarity of notation, we denote the follower's decision and his/her fund with herd coefficient $\eta$ as $\boldsymbol{\pi}(\eta):=\{\pi_t(\eta)\}_{t\in\mathscr{T}}$ and $\boldsymbol{x}(\eta):=\{x_t(\eta)\}_{t\in\mathscr{T}}$, respectively.

\subsection{The Regulator}
The regulator aims to maximize social welfare and minimize regulatory cost. In the simple scenario with two investors considered in this work, maximizing the social welfare is equivalent to maximizing the follower's expected terminal fund $\mathbb{E}x_T(\eta)$, as the leader's expected terminal fund is fixed given his/her decision in \eqref{equ:leader}. To achieve this, the regulator needs to design a mechanism to adjust the follower's level of herding, encouraging the follower to allocate more funds to the risky asset.

\subsubsection{The Policy}
Given the follower's herd coefficient $\eta$, the regulator implements a policy $q(\eta)$, which adjusts the follower's herd coefficient by an amount $u(q(\eta))$. We focus on the case where the regulator aims to reduce $\eta$, thereby mitigating excessive herding, rather than increasing it (\cite{bikhchandani1992theory}). For instance, $q(\eta)$ may represent the extent to which the regulator supplements the leader's decision with rational recommendations, e.g., mandating additional disclosure of information or issuing counter-herding advisories. A higher value of $q(\eta)$ provides the follower with stronger rational guidance, with $u(q(\eta))$ measuring the resulting reduction in the follower's level of herding (\cite{bikhchandani1992theory}). We assume that the policy does not alter the leader's decision, as leaders are typically rational investors (\cite{pratt1978risk}). Meanwhile, the policy implementation imposes a regulatory cost on the regulator. We define the regulatory cost as a function of the policy, denoted by $w(q(\eta))$, which captures the resources and effort required to introduce recommendations.

\subsubsection{The Compensation}
Under the regulator's policy, the follower's objective function is distorted and does not align with \eqref{equ:without_regulation}. To ensure that the follower accepts the regulation, the regulator must provide compensation alongside the policy implementation. Specifically, given the follower's herd coefficient $\eta$, the regulator provides a compensation $c(\eta)$, which enhances the follower's objective function by an amount $v(c(\eta))$ (\cite{hu2023inertia}). For instance, $c(\eta)$ may include fund incentives such as subsidies and transaction fee reductions, with $v(c(\eta))$ quantifying the follower's resulting utility gain (\cite{abdulsalam2021individualized}). When the regulator provides compensation, it constitutes a transfer from the regulator to the follower: the regulator's fund decreases while the follower's fund increases. We assume these two effects are approximately equal, so the transfer does not alter social welfare or add extra regulatory cost. In future work, we will relax this assumption and consider asymmetric transfers, where compensation introduces an extra cost term.

\subsubsection{The Reporting function}
Due to asymmetric information, the true value of the follower's herd coefficient $\eta$ is often unknown to the regulator, so the regulator requires the follower to report its value. We denote the reported herd coefficient as $\hat{\eta}:=s(\eta)$, where $s\in\mathscr{S}:\mathbb{R}_+\to\mathbb{R}_+$ is the follower's \textit{{reporting function}}. We assume that the follower tends to report a value less than the true herd coefficient, i.e., $\hat{\eta}\in[0,\eta]$, as overstating it may lead to a stricter regulation and reduce utility (\cite{myerson1981optimal}).

\subsection{The Overall Optimization Problem}
We refer to $q\in\mathscr{Q}:\mathbb{R}_+\to\mathbb{R}_+$ as the regulator's \textit{{policy}}, $c\in\mathscr{C}:\mathbb{R}_+\to\mathbb{R}_+$ as his/her \textit{{compensation}}, and define the pair $\gamma:=(q,c)\in\mathscr{G}$ as his/her \textit{{mechanism}}, where $\mathscr{G}:=\mathscr{Q}\times\mathscr{C}$ is the Cartesian product. We refer to $u:\mathbb{R}_+\to\mathbb{R}_+$ as the follower's \textit{{policy utility}}, $v:\mathbb{R}_+\to\mathbb{R}_+$ as his/her \textit{{compensation utility}}, and $w:\mathbb{R}_+\to\mathbb{R}_+$ as the regulator's \textit{{cost function}}. Following the work in \cite{borgers2015introduction}, we assume that $u$ and $v$ are common knowledge to both the regulator and the follower. We assume that the policy utility $u$ and the compensation utility $v$ increase with the policy $q$ and the compensation $c$. That is, a stronger policy leads to a greater reduction in the follower's herd coefficient, and a higher compensation results in a greater utility gain. We assume that $u(0)=v(0)=0$, i.e., the absence of policy or compensation yields no reduction in the herd coefficient and no utility gain. Additionally, we assume that $u(q(\eta))<q(\eta)$ for all $\eta\in\mathbb{R}_+$, i.e., the reduction in the follower's herd coefficient induced by the policy is less than the magnitude of the policy itself. From the work in \cite{dia2018fixed}, the regulator's policy $q(\eta)$ typically incurs a constant cost $\kappa\in\mathbb{R}_+$, since introducing recommendations often requires a baseline amount of resources and effort regardless of its intensity. We model the cost function as $w(q(\eta)):=\kappa\cdot\theta(q(\eta))$, where $\theta(x)$ is the unit step function. 
We assume the follower does not know the value of the parameter $\kappa$ and believes there is no regulatory cost when making his/her decisions.

In summary, the follower's optimization variables are the decision $\boldsymbol{\pi}(\eta)$ and the reporting function $s(\eta)$, while the regulator optimizes the mechanism $\gamma(\hat{\eta})$. This resembles a \textit{{Stackelberg game}} (\cite{shao2023risk,lv2023linear}), where the regulator acts first, and the follower responds subsequently. The overall optimization problem is
\begin{equation*}
    \left\{\begin{aligned}
        (\mathsf{P_1})\!\!\!\!\!\!\!\!&&\sup_{\gamma\in\mathscr{G}}\quad\!\!\!\!&\mathbb{E}x_T(\hat{\eta})-w(q(\hat{\eta})),\\
        (\mathsf{P_2})\!\!\!\!\!\!\!\!&&\sup_{\boldsymbol{\pi}\in\mathscr{P},s\in\mathscr{S}}\!\!\!\!&\mathbb{E}\varphi(x_T(\eta))-(\eta-u(q(\hat{\eta})))\delta(\boldsymbol{\pi}(\eta))+v(c(\hat{\eta})),
    \end{aligned}\right.
\end{equation*}
subject to \eqref{equ:fund_process}, where $\mathsf{P_1}$ and $\mathsf{P_2}$ are the regulator's and the follower's optimization problems, respectively.

\section{The Optimal Decision and Mechanism}
\label{sec:solution}
Next, we solve the optimal regulation problem $\mathsf{P_1}$ to find the regulator's optimal mechanism $\gamma^*(\hat{\eta}):=(q^*(\hat{\eta}),c^*(\hat{\eta}))$, and solve the optimal investment problem $\mathsf{P_2}$ to find the follower's optimal decision $\boldsymbol{\pi}^*(\eta)$ and optimal reporting function $s^*(\eta)$. 

The \textit{{revelation principle}} in \cite{myerson1979incentive} shows that, without loss of generality, the search for the optimal mechanism $\gamma^*(\hat{\eta})$ can be restricted to those under which the follower truthfully reports his/her herd coefficient $\eta$ subject to the IR and IC constraints. Thus, the follower's optimal reporting function $s^*(\eta)$ is an identity function, i.e., $s^*(\eta)=\eta$, and such mechanisms $\gamma^*(\eta)$ are referred to as truth-telling mechanisms. Below, we focus on the truth-telling mechanisms and let $s^*(\eta)=\eta$. 

Following the general approach in solving Stackelberg games, we proceed in three steps. First, we find the follower's optimal response $\boldsymbol{\omega}(\eta):=\{\omega_t(\eta)\}_{t\in\mathscr{T}}$ given the regulator's mechanism $\gamma(\eta)$. Next, we find the regulator's optimal mechanism $\gamma^*(\eta)$ given the follower's optimal response $\boldsymbol{\omega}(\eta)$. Then, we substitute the regulator's optimal mechanism $\gamma^*(\eta)$ into the follower's optimal response $\boldsymbol{\omega}(\eta)$ to obtain the follower's optimal decision $\boldsymbol{\pi}^*(\eta)$.

\subsection{The Follower's Optimal Response $\boldsymbol{\omega}(\eta)$ given $\gamma(\eta)$}

\label{subsec:opt_response}
From the work in \cite{wang2024investment}, given the mechanism $\gamma(\eta)$, the follower's optimal response and expected terminal fund are as in Theorem \ref{thm:opt_response}.

\begin{thm}\label{thm:opt_response}
    Given the regulator's mechanism $\gamma(\eta)$, the follower's optimal response is
    \begin{equation}\label{equ:varpi}
        \omega_t(\eta)=\frac{\tilde{\alpha}\sigma^2\mu(\eta)+\eta-u(q(\eta))}{\alpha\sigma^2\mu(\eta)+\eta-u(q(\eta))}\cdot\tilde{\pi}_t,
    \end{equation}
    where $\mu(\eta)$ is the integral parameter satisfying
    \begin{equation}\label{equ:mu}
        \mu(\eta)=\exp\left\{\left(\frac{(\alpha/\tilde{\alpha}-1)^2(\eta-u(q(\eta)))^2}{(\alpha\sigma^2\mu(\eta)+\eta-u(q(\eta)))^2}-1\right)\frac{\nu^2T}{2\sigma^2}\right\},
    \end{equation}
    and the follower's expected terminal fund is
    \begin{equation}\label{equ:ExT}
        \mathbb{E}x_T(\eta)=\frac{\tilde{\alpha}\sigma^2\mu(\eta)+\eta-u(q(\eta))}{\alpha\sigma^2\mu(\eta)+\eta-u(q(\eta))}\cdot\frac{\nu^2T}{\tilde{\alpha}\sigma^2}.
    \end{equation}
\end{thm}

\subsection{The Regulator's Optimal Mechanism $\gamma^*(\eta)$}
\label{subsec:opt_mechanism}

In mechanism design theory, an optimal mechanism typically satisfies both the IR constraint, i.e., followers are willing to accept the mechanism, and the IC constraint, i.e., followers truthfully report their private information. Next, following the general approach in mechanism design, we study the IR and IC constraints of the problems $\mathsf{P_1}$ and $\mathsf{P_2}$, and then solve for the regulator's optimal mechanism $\gamma^*(\eta)$ given the follower's optimal response $\boldsymbol{\omega}(\eta)$.

\subsubsection{The IR Constraint}
The IR constraint indicates that the objective function corresponding to the follower's optimal response with regulation in the problem $\mathsf{P_2}$ should be no less than that without regulation in \eqref{equ:without_regulation}. From the work in \cite{wang2024investment}, the follower's optimal decision without regulation $\bar{\boldsymbol{\pi}}(\eta):=\{\bar{\pi}_t(\eta)\}_{t\in\mathscr{T}}$ is
\begin{equation}\label{equ:barpi}
    \bar{\pi}_t(\eta)=\frac{\tilde{\alpha}\sigma^2\bar{\mu}(\eta)+\eta}{\alpha\sigma^2\bar{\mu}(\eta)+\eta}\cdot\tilde{\pi}_t,
\end{equation}
where the integral parameter $\bar{\mu}(\eta)$ satisfies
\begin{equation}\label{equ:barmu}
    \bar{\mu}(\eta)=\exp\left\{\left(\frac{(\alpha/\tilde{\alpha}-1)^2\eta^2}{(\alpha\sigma^2\bar{\mu}(\eta)+\eta)^2}-1\right)\frac{\nu^2T}{2\sigma^2}\right\},
\end{equation}
and the corresponding expected terminal fund is
\begin{equation}\label{equ:barExT}
    \mathbb{E}\bar{x}_T(\eta)=\frac{\tilde{\alpha}\sigma^2\bar{\mu}(\eta)+\eta}{\alpha\sigma^2\bar{\mu}(\eta)+\eta}\cdot\frac{\nu^2T}{\tilde{\alpha}\sigma^2}.
\end{equation}
We denote the objective functions concerning the follower's optimal response with and without regulation as $\mathbb{E}\varphi(x_T(\eta))$ and $\mathbb{E}\varphi(\bar{x}_T(\eta))$, respectively. Given the above notations, the IR constraint can be expressed as
\begin{equation}
    \mathbb{E}\varphi(x_T(\eta))\hspace{-0.25em}-\hspace{-0.25em}\eta \delta(\boldsymbol{\omega}(\eta))\hspace{-0.25em}+\hspace{-0.25em}v(c(\eta))\hspace{-0.25em}\geqslant\hspace{-0.25em}\mathbb{E}\varphi(\bar{x}_T(\eta))\hspace{-0.25em}-\hspace{-0.25em}\eta \delta(\bar{\boldsymbol{\pi}}(\eta)).\label{equ:IR_inequ}
\end{equation}

\begin{thm}\label{thm:IR}
    The IR constraint in \eqref{equ:IR_inequ} is equivalent to
    \begin{equation}
        c(\eta)\geqslant v^{-1}(f(\eta)),\label{equ:IR}
    \end{equation}
    where we denote
    \begin{equation}\label{equ:f}
        f(\eta):=\alpha^{-1}(\mu(\eta)-\bar{\mu}(\eta))+\eta(\delta(\boldsymbol{\omega}(\eta))- \delta(\bar{\boldsymbol{\pi}}(\eta))).
    \end{equation}
\end{thm}

Theorem \ref{thm:IR} shows that to ensure that the follower voluntarily accepts regulation, the regulator's optimal compensation has a lower bound $v^{-1}(f(\eta))$.

\subsubsection{The IC Constraint}
The IC constraint indicates that the follower's objective function when truthfully reporting his/her herd coefficient $\eta$ should be no less than that when reporting an arbitrary value $\hat{\eta}\in[0,\eta]$. When the follower's true herd coefficient is $\eta$ and he/she reports $\hat{\eta}$, we denote his/her optimal response as $\hat{\boldsymbol{\omega}}(\eta,\hat{\eta}):=\{\hat{\omega}_t(\eta,\hat{\eta})\}_{t\in\mathscr{T}}$. From \eqref{equ:varpi} and \eqref{equ:mu}, we have
\begin{equation}\label{equ:response}
    \hat{\omega}_t(\eta,\hat{\eta}):=\frac{\tilde{\alpha}\sigma^2\hat{\mu}(\eta,\hat{\eta})+\eta-u(q(\hat{\eta}))}{\alpha\sigma^2\hat{\mu}(\eta,\hat{\eta})+\eta-u(q(\hat{\eta}))}\cdot\tilde{\pi}_t,
\end{equation}
where the integral parameter $\hat{\mu}(\eta,\hat{\eta})$ satisfies
\begin{equation}\label{equ:hat_mu}
    \hat{\mu}(\eta,\hat{\eta})=\exp\left\{\left(\frac{(\alpha/\tilde{\alpha}-1)^2(\eta-u(q(\hat{\eta})))^2}{(\alpha\sigma^2\hat{\mu}(\eta,\hat{\eta})+\eta-u(q(\hat{\eta})))^2}-1\right)\frac{\nu^2T}{2\sigma^2}\right\},
\end{equation}
and we denote the fund under $\hat{\boldsymbol{\omega}}(\eta,\hat{\eta})$ as $\hat{\boldsymbol{x}}(\eta,\hat{\eta}):=\{\hat{x}_t(\eta,\hat{\eta})\}_{t\in\mathscr{T}}$. When the follower truthfully reports his/her herd coefficient, i.e., $\hat{\eta}=\eta$, we have $\hat{\boldsymbol{\omega}}(\eta,\eta)=\boldsymbol{\omega}(\eta)$, $\hat{\mu}(\eta,\eta)=\mu(\eta)$, and $\hat{\boldsymbol{x}}(\eta,\eta)=\boldsymbol{x}(\eta)$. Given the above notations, the IC constraint can be expressed as
\begin{align}
    &\mathbb{E}\varphi(\hat{x}_T(\eta,\eta))-\eta \delta(\hat{\boldsymbol{\omega}}(\eta,\eta))+v(c(\eta))\notag\\
    \geqslant\ &\mathbb{E}\varphi(\hat{x}_T(\eta,\hat{\eta}))-\eta \delta(\hat{\boldsymbol{\omega}}(\eta,\hat{\eta}))+v(c(\hat{\eta})).\label{equ:IC_inequ}
\end{align}
\begin{thm}\label{thm:IC}
    The IC constraint in \eqref{equ:IC_inequ} is equivalent to
    \begin{equation}\label{equ:IC}
        c(\eta)=v^{-1}\left(\int_0^\eta u(q(\xi))\psi(\xi)\dif\xi+\chi\right),
    \end{equation}
    where we denote $\psi(\eta):=\left.\frac{\partial \delta(\hat{\boldsymbol{\omega}}(\eta,\hat{\eta}))}{\partial\hat{\eta}}\right|_{\hat{\eta}=\eta}$, and $\chi\in\mathbb{R}_+$ is a parameter determined by the IR constraint and policy.
\end{thm}

Theorem \ref{thm:IC} shows that the marginal compensation utility $v'(c(\eta))$ equals the product of the policy $u(q(\eta))$ and the marginal deviation concerning the reported herd coefficient. Intuitively, this means that the compensation is set to exactly offset the impact of the policy on the follower's objective function, which ensures that the follower has no incentive to misreport his/her true herd coefficient. Additionally, to satisfy the IR constraint, from \eqref{equ:IC}, a nonnegative compensation $v^{-1}(\chi)$ is required to ensure that the follower accepts the regulation. Note that $\chi$ does not vary with $\eta$, meaning that, regardless of the follower's level of herding, the regulator must offer this compensation when imposing policy. We refer to $\chi$ as the \textit{{constant compensation utility}} and will derive its expression later.

\subsubsection{The Optimal Mechanism}
Next, we derive the regulator's optimal mechanism $\gamma^*(\eta)$. Note that under the IC constraint in \eqref{equ:IC}, the optimal compensation $c^*(\eta)$ is a function of the optimal policy $q^*(\eta)$, we first solve for the optimal policy $q^*(\eta)$, and then determine the optimal compensation $c^*(\eta)$ given $q^*(\eta)$.

\uline{{The optimal policy $q^*(\eta)$.}} We first consider the simple case without regulatory cost, i.e., $\kappa=0$, and then extend to the general case with a positive regulatory cost, i.e., $\kappa>0$. 

Case 1 ($\kappa=0$). The optimal policy is as in Theorem \ref{thm:opt_policy_circ}.

\begin{thm}\label{thm:opt_policy_circ}
    If $\kappa=0$, the regulator's optimal policy is
    \begin{equation}\label{equ:opt_policy_circ}
        q^*(\eta)=\eta\cdot\theta(\tilde{\alpha}-\alpha).
    \end{equation}
\end{thm}

Theorem \ref{thm:opt_policy_circ} shows that without the regulatory cost, the regulator's optimal policy exhibits a switch-like structure. If the follower is more risk-averse than the leader, the follower allocates fewer funds to the risky asset without considering the leader's influence. In this case, herding causes the follower's decision to align closer to the leader's decision, promoting a higher allocation to the risky asset and increasing the social welfare. Thus, the regulator has no incentive to reduce the follower's herd coefficient. If the follower is more risk-taking than the leader, herding causes the follower to allocate fewer funds to the risky asset and thus lowers the social welfare, and the regulator imposes the strongest policy to reduce the herd coefficient.

Furthermore, from Theorem \ref{thm:opt_policy_circ} and \eqref{equ:varpi}--\eqref{equ:ExT}, if $\kappa=0$, the follower's optimal decision is
\begin{equation}\label{equ:pi*}
    \pi_t^*(\eta)=\frac{\tilde{\alpha}\sigma^2\mu^*(\eta)+\eta-u(\eta)}{\alpha\sigma^2\mu^*(\eta)+\eta-u(\eta)}\cdot\tilde{\pi}_t,
\end{equation}
where the integral parameter satisfies
\begin{equation}\label{equ:mu*}
    \mu^*(\eta)=\exp\left\{\left(\frac{(\alpha/\tilde{\alpha}-1)^2(\eta-u(\eta))^2}{(\alpha\sigma^2\mu^*(\eta)+\eta-u(\eta))^2}-1\right)\frac{\nu^2T}{2\sigma^2}\right\},
\end{equation}
and the corresponding expected terminal fund is
\begin{equation}\label{equ:ExT*}
    \mathbb{E}x_T^*(\eta)=\frac{\tilde{\alpha}\sigma^2\mu^*(\eta)+\eta-u(\eta)}{\alpha\sigma^2\mu^*(\eta)+\eta-u(\eta)}\cdot\frac{\nu^2T}{\tilde{\alpha}\sigma^2}.
\end{equation}

Case 2 ($\kappa>0$). Next, we study the general case with a positive regulatory cost. Note that the difference in the expected terminal funds between the regulated and unregulated scenarios is finite, as the follower's decision is assumed to be a measurable and absolutely integrable process. We define this difference as the \textit{{economic gain}}, denoted by $g(\eta):=\mathbb{E}x_T^*(\eta)-\mathbb{E}\bar{x}_T(\eta)$. If the regulatory cost $\kappa$ exceeds the supremum of the economic gain $g(\eta)$ for all $\eta\in\mathbb{R}_+$, the regulator will not implement the policy, because the intervention always incurs a net loss in terms of social welfare. Thus, we only study the case when the regulatory cost $\kappa \in \mathscr{K}:=[0, \sup_{\eta\in\mathbb{R}_+}g(\eta)]$.
\begin{thm}\label{thm:opt_policy}
    If $\kappa\in\mathscr{K}$, the regulator's optimal policy is
    \begin{equation}\label{equ:opt_policy}
        q^*(\eta)=\eta\cdot\theta(\tilde{\alpha}-\alpha)\cdot\theta(\eta-\breve{\eta}),
    \end{equation}
    where we denote $\breve{\eta}:=g^{-1}(\kappa)$.
\end{thm}

Theorem \ref{thm:opt_policy} shows that with the regulatory cost, the optimal policy still exhibits a switch-like structure as in the case without the regulatory cost. The difference is that without the regulatory cost, the optimal policy depends solely on the risk coefficients of the leader and the follower, whereas with the regulatory cost, the regulator also weighs the economic gain against the regulatory cost. This tradeoff results in the herd coefficient threshold, and the policy is imposed only when the follower's herd coefficient exceeds this threshold. The regulatory cost makes the regulator intervene selectively, applying the policy only to followers whose herding is strong. Otherwise, intervention will result in a net loss and should be avoided. In summary, from Theorems \ref{thm:opt_policy_circ} and \ref{thm:opt_policy}, the regulator intervenes only when the follower is more risk-taking than the leader and exhibits strong herding, and imposes no regulation otherwise.

Next, we examine the impact of the constant regulatory cost $\kappa$ on the threshold $\breve{\eta}$. We focus on the case where the follower is more risk-taking than the leader, as the optimal policy in \eqref{equ:opt_policy} does not involve the threshold when $\alpha\geqslant\tilde{\alpha}$.

\begin{thm}\label{thm:threshold}
    If $\alpha<\tilde{\alpha}$, the herd coefficient threshold $\breve{\eta}$ is an increasing function of the constant regulatory cost $\kappa$.
\end{thm}

Theorem \ref{thm:threshold} shows that as the constant regulatory cost increases, the threshold also increases. Intuitively, this means that as the constant regulatory cost increases, the threshold also increases, reflecting that regulators must focus on the followers whose herding is strong enough to ensure that the economic gain exceeds the regulatory cost.

\uline{{The optimal compensation $c^*(\eta)$.}} Next, we derive the regulator's optimal compensation in Theorem \ref{thm:opt_compensation}.

\begin{thm}\label{thm:opt_compensation}
    The regulator's optimal compensation is
    \begin{equation}\label{equ:opt_compensation}
        c^*(\eta)=v^{-1}\left(\theta(\tilde{\alpha}-\alpha)\cdot\theta(\eta-\breve{\eta})\cdot\int_{\breve{\eta}}^{\eta} u(\xi)\psi(\xi)\dif\xi+\chi\right),
    \end{equation}
    where the constant compensation utility satisfies
    \begin{equation}
        \chi=\int_{\mathbb{R}_+}\delta(\boldsymbol{\omega}(\eta))-\delta(\bar{\boldsymbol{\pi}}(\eta))-u(\eta)\phi(\eta)\dif\eta,\label{equ:necessary}
    \end{equation}
    and we denote $\phi(\eta):=\left.\frac{\partial \delta(\hat{\boldsymbol{\omega}}(\eta,\hat{\eta}))}{\partial\eta}\right|_{\hat{\eta}=\eta}$.
\end{thm}

Theorem \ref{thm:opt_compensation} shows that when the follower is more risk-averse than the leader, or when the follower's herd coefficient is below the threshold, no compensation is required, since in this case, the optimal policy is zero. If the follower is more risk-taking than the leader and his/her herd coefficient exceeds the threshold, the regulator imposes the strongest policy to mitigate herding, enabling a higher expected terminal fund. Thus, the optimal compensation serves to cover the follower's utility loss due to regulatory intervention, ensuring individual rationality and incentive compatibility. The optimal compensation $c^*(\eta)$ in \eqref{equ:opt_compensation} consists of two components. The first component is the constant compensation utility $\chi$, which is independent of the herding coefficient $\eta$ and ensures the IR constraint. The second component increases with the follower's herding coefficient $\eta$ to ensure the IC constraint. Additionally, both the optimal policy and the $\eta$-dependent component of optimal compensation increase with the follower's herd coefficient $\eta$, indicating that more pronounced herding triggers a stronger policy and higher compensation. 

\subsection{The Follower's Optimal Decision $\boldsymbol{\pi}^*(\eta)$}
\label{subsec:opt_decision}

\begin{thm}\label{thm:opt_decision}
    The follower's optimal decision is
    \begin{equation}\label{equ:opt_decision}
        \pi^*_t(\eta)=
        \begin{cases}
            \displaystyle\frac{\tilde{\alpha}\sigma^2\bar{\mu}(\eta)+\eta}{\alpha\sigma^2\bar{\mu}(\eta)+\eta}\cdot\tilde{\pi}_t,&\alpha\geqslant\tilde{\alpha}\ \text{or}\ \eta\leqslant\breve{\eta}\\
            \displaystyle\frac{\tilde{\alpha}\sigma^2\mu^*(\eta)+\eta-u(\eta)}{\alpha\sigma^2\mu^*(\eta)+\eta-u(\eta)}\cdot\tilde{\pi}_t,&\alpha<\tilde{\alpha}\ \text{and}\ \eta>\breve{\eta}
        \end{cases}.
    \end{equation}
\end{thm}

Theorem \ref{thm:opt_decision} shows that if the follower is more risk-averse than the leader or the follower's herd coefficient is below the threshold, the follower's optimal decision coincides with the optimal decision without regulation, since no policy is imposed. If the follower is more risk-taking than the leader and the follower's herd coefficient exceeds the threshold, the regulator reduces the follower's herd coefficient, pushing the follower's decision away from the leader's more risk-averse decision, resulting in a higher optimal decision than that without regulation.

\section{\!Impact of Regulation on Social Welfare\!}
\label{sec:analysis}
In this section, we study the impact of regulation on social welfare. We employ the economic gain to capture the improvement in social welfare through the mechanism.
\begin{thm}\label{thm:reg_gain}
    For all $\eta\in\mathbb{R}_+$, the economic gain satisfies that $g(\eta)=0$ if $\alpha\geqslant\tilde{\alpha}$ or $\eta\leqslant\breve{\eta}$, while $g(\eta)\geqslant0$ and $\dif g(\eta)/\dif\eta\geqslant0$ if $\alpha<\tilde{\alpha}$ and $\eta>\breve{\eta}$.
\end{thm}

Theorem \ref{thm:reg_gain} shows that if the follower is more risk-averse than the leader or the follower's herd coefficient is below the threshold, the social welfare is the same under both the regulated and unregulated scenarios. In this case, the social welfare has already reached its upper bound, and the regulator is unable to improve it further through intervention. If the follower is more risk-taking than the leader and the follower's herd coefficient exceeds the threshold, the regulator mitigates the follower's herding, leading to a higher level of social welfare compared to the unregulated scenario. Additionally, as herding becomes more pronounced, regulation leads to a more substantial improvement in social welfare than the unregulated scenario.

\section{Conclusion}
\label{sec:conclusion}
In this work, we propose a regulator-leader-follower trilateral game framework that integrates optimal investment theory with mechanism design to study regulatory interventions under herding. We analyze the complex dynamics among the regulator, the leader, and the follower, and design the optimal regulatory mechanisms to effectively mitigate herding and theoretically analyze its influence on social welfare. Our analysis shows that the optimal policy exhibits a switch-like structure. When the follower is more risk-averse than the leader, no intervention is required. When the follower is more risk-taking, regulation is imposed if the herd coefficient exceeds a threshold, and the regulator imposes the strongest policy and provides compensation to ensure incentive compatibility. Furthermore, the economic gain increases with the level of herding. 

\bibliography{ifacconf}

\clearpage

\onecolumn

\appendix

\vspace*{12pt}
\begin{center}
    {\fontsize{14}{16}\selectfont\bfseries The Supplementary File} \\[8pt]
    {\normalsize\bfseries Huisheng Wang \quad H. Vicky Zhao}
\end{center}
\vspace{12pt}

\section{Proof of Theorems}
\subsection{Theorem \ref{thm:IR}}
From the work in \cite{wang2024investment}, we have
\begin{equation}
    \mathbb{E}\varphi(x_T(\eta))=-\alpha^{-1}\mu(\eta)\ \text{and}\ \mathbb{E}\varphi(\bar{x}_T(\eta))=-\alpha^{-1}\bar{\mu}(\eta).
\end{equation}
Since $v(c(\eta))$ increases with $c(\eta)$, $v^{-1}(f(\eta))$ also increases with $f(\eta)$. Therefore, from \eqref{equ:IR_inequ}, we can obtain \eqref{equ:IR}.

\subsection{Theorem \ref{thm:IC}}
From \eqref{equ:IC_inequ}, we have
\begin{equation}\label{equ:IC_condition}
    \frac{\dif v(c(\eta))}{\dif \eta}+\left.\frac{\partial\mathbb{E}\varphi(\hat{x}_T(\eta,\hat{\eta}))}{\partial\hat{\omega}_t(\eta,\hat{\eta})}\cdot\frac{\partial\hat{\omega}_t(\eta,\hat{\eta})}{\partial\hat{\eta}}\right|_{\hat{\eta}=\eta}-\left.\eta\frac{\partial \delta(\hat{\boldsymbol{\omega}}(\eta,\hat{\eta}))}{\partial\hat{\omega}_t(\eta,\hat{\eta})}\cdot\frac{\partial\hat{\omega}_t(\eta,\hat{\eta})}{\partial\hat{\eta}}\right|_{\hat{\eta}=\eta}=0.
\end{equation}
Additionally, the first-order optimality condition of the problem $\mathsf{P_2}$ is
\begin{equation}\label{equ:first_order_condition}
    \frac{\partial\mathbb{E}\varphi(\hat{x}_T(\eta,\hat{\eta}))}{\partial\hat{\omega}_t(\eta,\hat{\eta})}-(\eta-u(q(\hat{\eta})))\frac{\partial \delta(\hat{\boldsymbol{\omega}}(\eta,\hat{\eta}))}{\partial\hat{\omega}_t(\eta,\hat{\eta})}=0.
\end{equation}
Substituting \eqref{equ:first_order_condition} into \eqref{equ:IC_condition}, we have
\begin{equation}\label{equ:v'c}
    \frac{\dif v(c(\eta))}{\dif \eta}=\left.u(q(\eta))\frac{\partial \delta(\hat{\boldsymbol{\omega}}(\eta,\hat{\eta}))}{\partial\hat{\eta}}\right|_{\hat{\eta}=\eta}.
\end{equation}
Since $u(0)=v(0)=0$, from \eqref{equ:v'c}, we obtain \eqref{equ:IC}.

\subsection{Theorem \ref{thm:opt_policy_circ}}
If $\alpha\geqslant\tilde{\alpha}$, in the following, we prove that $\mathbb{E}x_T(\eta)$ decreases with the policy $q(\eta)$. That is, a stronger policy mitigates herding, leads to a lower allocation to the risky asset, and reduces the social welfare. The optimal policy is $q^*(\eta)=0$ for all $\eta\in\mathbb{R}_+$, i.e., no policy is imposed on the follower's herd coefficient. If $\alpha<\tilde{\alpha}$, in the following, we prove that $\mathbb{E}x_T(\eta)$ increases with the policy $q(\eta)$. Thus, the optimal policy $q^*(\eta)=\eta$, i.e., the regulator imposes the strongest policy to reduce the follower's herd coefficient.

From \eqref{equ:mu}, we have
\begin{equation}
    \frac{\dif \mu(\eta)}{\dif u(q(\eta))} = -\frac{ \left(\alpha/\tilde{\alpha} - 1\right)^2 \mu(\eta) \frac{2\alpha \sigma^2 \mu(\eta)(\eta - u(q(\eta)))} {\left( \alpha \sigma^2 \mu(\eta) + \eta - u(q(\eta)) \right)^3} \cdot \frac{\nu^2 T}{2\sigma^2} } { 1 + \left(\alpha/\tilde{\alpha} - 1\right)^2 \mu(\eta) \frac{2\alpha \sigma^2(\eta - u(q(\eta)))^2} {\left( \alpha \sigma^2 \mu(\eta) + \eta - u(q(\eta)) \right)^3} \cdot \frac{\nu^2 T}{2\sigma^2}}<0.
\end{equation}
That is, the integral parameter $\mu(\eta)$ is a strictly decreasing function of the policy utility $u(q(\eta))$. Given that $\frac{\dif u(q(\eta))}{\dif q(\eta)}\geqslant0$, i.e., the policy utility increases with the policy, it follows that 
\begin{equation}\label{equ:mu_q}
    \frac{\dif \mu(\eta)}{\dif q(\eta)} =\frac{\dif u(q(\eta))}{\dif q(\eta)}\cdot\frac{\dif \mu(\eta)}{\dif u(q(\eta))}\leqslant0.
\end{equation}
That is, the integral parameter $\mu(\eta)$ is a decreasing function of the policy $q(\eta)$. To simplify the notation, we define
\begin{equation}\label{equ:z}
    z(q(\eta)):=\frac{\mu(\eta)}{\eta-u(q(\eta))}\in\mathbb{R}_+.
\end{equation}
From \eqref{equ:mu}, we have
\begin{equation}
    \mu(\eta)=\exp\left\{\left(\frac{(\alpha/\tilde{\alpha}-1)^2}{\left(\alpha\sigma^2z(q(\eta))+1\right)^2}-1\right)\frac{\nu^2T}{2\sigma^2}\right\}.
\end{equation}
Therefore, we have
\begin{equation}
    z(q(\eta))= \frac{1}{\alpha \sigma^2} \left( \sqrt{ \frac{ \left( \alpha/\tilde{\alpha} - 1 \right)^2 }{ 1 + \frac{2 \sigma^2}{\nu^2 T} \ln \mu(\eta) } } - 1 \right).\label{equ:z_mu}
\end{equation}
From \eqref{equ:z_mu}, we can prove that $z(q(\eta))$ is an increasing function of the integral parameter $\mu(\eta)$. From \eqref{equ:mu_q}, since the integral parameter $\mu(\eta)$ is a decreasing function of the policy $q(\eta)$, $z(q(\eta))$ is an increasing function of the policy $q(\eta)$, i.e., $\frac{\dif z(q(\eta))}{\dif q(\eta)}\geqslant0$. From \eqref{equ:ExT}, we have
\begin{equation}
    \frac{\dif \mathbb{E}x_T(\eta)}{\dif q(\eta)}=-\frac{\dif z(q(\eta))}{\dif q(\eta)}\cdot\frac{(\alpha/\tilde{\alpha}-1)\nu^2 T}{\left( \alpha \sigma^2 z(q(\eta)) + 1 \right)^2}.
\end{equation}
Therefore, we can prove that $\frac{\dif \mathbb{E}x_T(\eta)}{\dif q(\eta)}\leqslant0$ when $\alpha\geqslant\tilde{\alpha}$ and $\frac{\dif \mathbb{E}x_T(\eta)}{\dif q(\eta)}\geqslant0$ when $\alpha<\tilde{\alpha}$. 

\subsection{Theorem \ref{thm:opt_policy}}
If $\alpha\geqslant\tilde{\alpha}$, from Theorem \ref{thm:opt_policy_circ}, we can prove that the optimal policy is $q^*(\eta)=0$ for all $\eta\in\mathbb{R}_+$ with the regulatory cost. 
If $\alpha<\tilde{\alpha}$, since the regulatory cost is a step function, the optimal policy $q^*(\eta)$ should be chosen to maximize the higher of the two: (i) the social welfare with regulation minus the constant regulatory cost, $\mathbb{E}x_T^*(\eta)-\kappa$, and (ii) the social welfare without regulation, $\mathbb{E}x_T(\eta)$. 
In the following, we prove that if the follower's herd coefficient $\eta$ exceeds the \textit{{threshold}} $\breve{\eta}$, the regulator's optimal policy is $q^*(\eta)=\eta$ and $q^*(\eta) = 0$ otherwise.
    
To prove that $\mathbb{E}x_T^*(\eta)-\kappa>\mathbb{E}x_T(\eta)$ if and only if $\eta>\breve{\eta}$, note that $g(\breve{\eta})=\kappa$. It therefore suffices to show that the economic gain $g(\eta)$ is a strictly increasing function of the herd coefficient $\eta$. This ensures that for all $\eta>\breve{\eta}$, $g(\eta)>\kappa$, i.e., $\mathbb{E}x_T^*(\eta)-\kappa>\mathbb{E}x_T(\eta)$, and conversely for $\eta<\breve{\eta}$. See the proof of Theorem \ref{thm:reg_gain}.

\subsection{Theorem \ref{thm:threshold}}
Since $\breve{\eta}$ is the inverse function of the economic gain given $\kappa$, we only need to prove the economic gain $g(\eta)$ increases with $\eta$. Theorem \ref{thm:reg_gain} shows that the economic gain $g(\eta)$ is an increasing function of the follower's herd coefficient $\eta$, which further implies that the threshold $\breve{\eta}$ is an increasing function of the constant regulatory cost $\kappa$.

\subsection{Theorem \ref{thm:opt_compensation}}
From \eqref{equ:opt_policy}, if $\alpha\geqslant\tilde{\alpha}$ or $\eta\leqslant\breve{\eta}$, the optimal policy is $q^*(\eta)=0$. The follower's optimal response with regulation coincides with the optimal decision without regulation, i.e., $\boldsymbol{\omega}(\eta)=\bar{\boldsymbol{\pi}}(\eta)$, and the corresponding integral constants are equal, i.e., $\mu(\eta)=\bar{\mu}(\eta)$. From the IR constraint in \eqref{equ:IR}, we have $c^*(\eta)\geqslant0$. Given that the regulator's compensation is as small as possible, his/her optimal compensation is $c^*(\eta)=v^{-1}(\chi)$ for all $\eta\in\mathbb{R}_+$, and it can be proved that the IC constraint in \eqref{equ:IC} is satisfied. 

If $\alpha<\tilde{\alpha}$ and $\eta>\breve{\eta}$, the optimal policy is $q^*(\eta)=\eta$. Since the follower believes there is no regulatory cost when making his/her decisions, from the IC constraint in \eqref{equ:IC}, we have the optimal compensation in \eqref{equ:opt_compensation}. To further ensure that it satisfies the IR constraint in \eqref{equ:IR}, we need to determine the constant compensation utility $\chi$. From \eqref{equ:IR} and \eqref{equ:IC}, since $v^{-1}(f(\eta))$ increases with $f(\eta)$, it can be proved that the constant compensation utility must satisfy
\begin{equation}
    \chi\geqslant \rho(\eta):=f(\eta)-\int_0^\eta u(\xi)\psi(\xi)\dif\xi,\label{equ:necessary_condition}
\end{equation}
for all $\eta\in\mathbb{R}_+$, where $\rho(\eta)$ represents the difference between the follower's compensation utility corresponding to the minimum compensation that satisfies the IR and IC conditions. From \eqref{equ:first_order_condition}, we can transform the the second term of $\rho(\eta)$ into
\begin{equation}\label{equ:int}
    \int_0^\eta u(\xi)\psi(\xi)\dif\xi=\underbrace{\int_0^\eta\left.-\frac{\partial\mathbb{E}\varphi(\hat{x}_T(\xi,\hat{\eta}))}{\partial\hat{\eta}}\right|_{\hat{\eta}=\xi}\dif\xi}_{:=I}+\underbrace{\int_0^\eta\xi\psi(\xi)\dif\xi}_{:=J}.
\end{equation}
Given $\hat{\boldsymbol{\omega}}(\eta,\eta)=\boldsymbol{\omega}(\eta)$, $\hat{\mu}(\eta,\eta)=\mu(\eta)$, and $\hat{\boldsymbol{x}}(\eta,\eta)=\boldsymbol{x}(\eta)$, from the relationship between the total derivative and partial derivatives, we have
\begin{align}
    \frac{\dif\delta(\boldsymbol{\omega}(\eta))}{\dif\eta}&=\psi(\eta)+\phi(\eta),\quad\text{and}\\
    \frac{\dif\mathbb{E}\varphi(x_T(\eta))}{\dif\eta}&=\left.\frac{\partial\mathbb{E}\varphi(\hat{x}_T(\eta,\hat{\eta}))}{\partial\eta}+\frac{\partial\mathbb{E}\varphi(\hat{x}_T(\eta,\hat{\eta}))}{\partial\hat{\eta}}\right|_{\hat{\eta}=\eta}.
\end{align}
Therefore, we can simplify the integrals $I$ and $J$ in \eqref{equ:int} as
\begin{align}
    I&=\alpha^{-1}(\mu(\eta)-\mu(0))+\int_0^\eta\left.\frac{\partial\mathbb{E}\varphi(\hat{x}_T(\xi,\hat{\eta}))}{\partial\xi}\right|_{\hat{\eta}=\xi}\dif\xi,\label{equ:I_}\\
    J&=\eta\delta(\boldsymbol{\omega}(\eta))-\int_0^\eta\delta(\boldsymbol{\omega}(\xi))\dif\xi-\int_0^\eta\xi\phi(\xi)\dif\xi.\label{equ:J_}
\end{align}
From \eqref{equ:first_order_condition}, we have 
\begin{equation}\label{equ:I0-J0}
    \left.\frac{\partial\mathbb{E}\varphi(\hat{x}_T(\eta,\hat{\eta}))}{\partial\eta}\right|_{\hat{\eta}=\eta}-\eta\phi(\eta)=-u(\eta)\phi(\eta).
\end{equation}
Combining \eqref{equ:I_}--\eqref{equ:I0-J0}, we can obtain
\begin{equation}\label{equ:upsi}
    \int_0^\eta u(\xi)\psi(\xi)\dif\xi=\alpha^{-1}(\mu(\eta)-\mu(0))+\eta\delta(\boldsymbol{\omega}(\eta))-\int_0^\eta\delta(\boldsymbol{\omega}(\xi))\dif\xi-\int_0^\eta u(\xi)\phi(\xi)\dif\xi.
\end{equation}
On the other hand, from \eqref{equ:without_regulation}, we have
\begin{equation}\label{equ:firstorder}
    \frac{\dif}{\dif\eta}(-\mathbb{E}\varphi(\bar{x}_T(\eta))+\eta\delta(\bar{\boldsymbol{\pi}}(\eta)))=\delta(\bar{\boldsymbol{\pi}}(\eta)).
\end{equation}
Therefore, by integrating both sides of \eqref{equ:firstorder}, we have
\begin{equation}\label{equ:inequ}
    \alpha^{-1}(\bar{\mu}(\eta)-\bar{\mu}(0))+\eta\delta(\bar{\boldsymbol{\pi}}(\eta))=\int_0^\eta\delta(\bar{\boldsymbol{\pi}}(\xi))\dif\xi.
\end{equation}
When the follower's herd coefficient $\eta=0$, the regulator's policy satisfies $q^*(\eta)=0$, and we have $\mu(0)=\bar{\mu}(0)$. Combining \eqref{equ:f}, \eqref{equ:upsi}, and \eqref{equ:inequ}, we have
\begin{equation}
    \rho(\eta)=\int_0^\eta\delta(\boldsymbol{\omega}(\xi))-\delta(\bar{\boldsymbol{\pi}}(\xi))+u(\xi)\phi(\xi)\dif\xi.\label{equ:diff}
\end{equation}
Since a larger herd coefficient implies a stronger herding that leads to a smaller change of deviation, we have $\phi(\eta)<0$. Additionally, given the same herd coefficient $\eta$, the deviation with regulation is greater than or equal to that without regulation, i.e., $\delta(\boldsymbol{\omega}(\eta)) \geqslant \delta(\bar{\boldsymbol{\pi}}(\eta))$. Since $u(q(\eta)) < q(\eta)$, from \eqref{equ:varpi} and \eqref{equ:barpi}, as $\eta$ approaches infinity, both deviations with regulation $\delta(\boldsymbol{\omega}(\eta))$ and without regulation $\delta(\bar{\boldsymbol{\pi}}(\eta))$ approach $0$. Additionally, from \eqref{equ:delta}, \eqref{equ:varpi}, and \eqref{equ:barpi}, we can prove that as $\eta$ approaches infinity, $\delta(\boldsymbol{\omega}(\eta)) = \mathrm{O}(\eta^{-2})$ and $\delta(\bar{\boldsymbol{\pi}}(\eta))=\mathrm{O}(\eta^{-2})$. Furthermore, from \eqref{equ:delta} and \eqref{equ:response}, we can prove that $u(\eta)\phi(\eta)=\mathrm{O}(\eta^{-3}u(\eta))$, which decays faster than $\eta^{-2}$ as $\eta$ approaches infinity. Therefore, the infinite integral $\lim_{\eta\to+\infty}\rho(\eta)$ converges, and $\rho(\eta)$ increases with $\eta$. From \eqref{equ:necessary_condition}, since the regulator minimizes compensation, we let equality hold and obtain \eqref{equ:necessary}.

\subsection{Theorem \ref{thm:reg_gain}}
If $\alpha\geqslant\tilde{\alpha}$ or $\eta\leqslant\breve{\eta}$, we have $\mathbb{E}x_T^*(\eta)=\mathbb{E}\bar{x}_T(\eta)$ and thus $g(\eta)=0$. If $\alpha<\tilde{\alpha}$ and $\eta>\breve{\eta}$, from \eqref{equ:z}, we have
\begin{equation}\label{equ:ExT_regulation_app}
    g(\eta)=\left(\frac{\tilde{\alpha}\sigma^2z(q(\eta))+1}{\alpha\sigma^2z(q(\eta))+1}-\frac{\tilde{\alpha}\sigma^2z(0)+1}{\alpha\sigma^2z(0)+1}\right)\frac{\nu^2T}{\tilde{\alpha}\sigma^2}.
\end{equation}
Since $\frac{\dif z(q(\eta))}{\dif q(\eta)}\geqslant0$, we have $z(q(\eta))\geqslant z(0)$. From \eqref{equ:ExT_regulation_app}, we have $g(\eta)\geqslant0$. Additionally, if $\alpha<\tilde{\alpha}$ and $\eta>\breve{\eta}$, we have
\begin{equation}
    \frac{\dif g(\eta)}{\dif\eta}=-\frac{\dif \mathbb{E}\bar{x}_T(\eta)}{\dif q(\eta)}\geqslant0.
\end{equation}
So far, we have finished the proof of Theorem \ref{thm:reg_gain}.

\section{Numerical Experiments}
\label{sec:experiment}
\vspace{-3mm}
In this section, we conduct numerical experiments to validate our analysis of how regulation influences the follower's decision and the social welfare. Following the work in \cite{wang2024investment}, we set the investment horizon $T=50$, the interest rate $r=0.04$, the excess return $\nu=0.03$, the volatility $\sigma=0.17$, the constant regulatory cost $\kappa\in\{0,0.5\}$, the leader's risk coefficient $\tilde{\alpha}=0.3$, the follower's risk coefficient $\alpha\in[0.2,0.35]$, and the herd coefficient $\eta\in[0,0.01]$, respectively. Additionally, the policy utility is set as $u(q(\eta))=0.9q(\eta)$ and the compensation utility is set as $v(c(\eta))=c(\eta)$.

\begin{figure}[H]
    \begin{center}
        \includegraphics[width=0.5\linewidth]{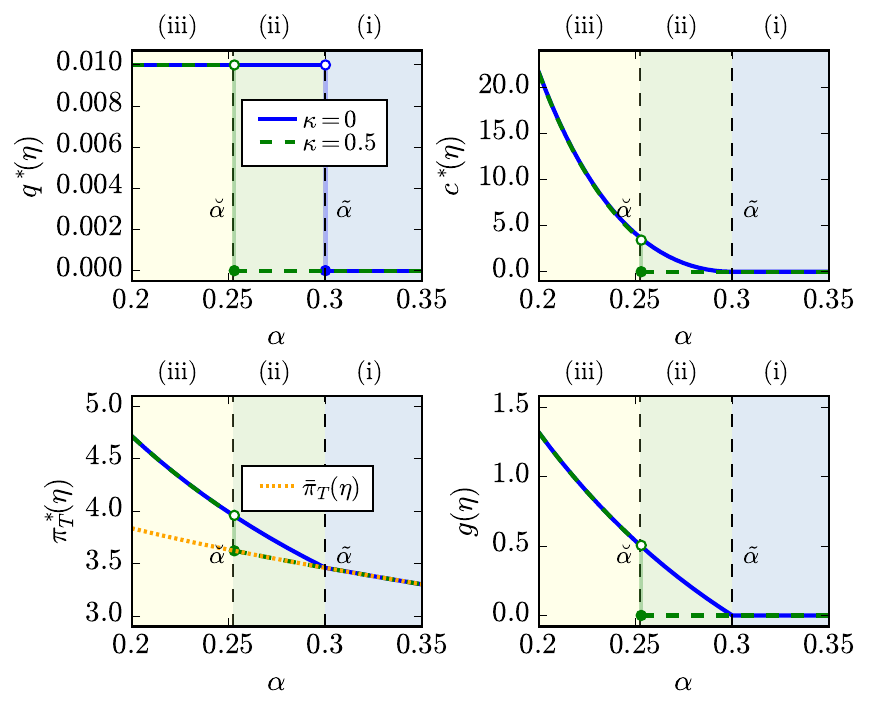}
        \vspace{-6mm}
        \caption{The regulator's optimal policy and compensation, the follower's optimal decision, and the economic gain when the follower's herd coefficient $\eta=0.01$.} 
        \label{fig:case1}
        \vspace{-1mm}
    \end{center}
\end{figure}

\subsection{Case 1 $({\eta=0.01})$}
First, we fix the follower's herd coefficient at $\eta=0.01$ and change the risk coefficient $\alpha$. We find that when the follower's risk coefficient $\alpha$ equals $\breve{\alpha}\approx0.252$, the herd coefficient threshold $\breve{\eta}$ equals the follower's herd coefficient $\eta$. If the follower's risk coefficient $\alpha<\breve{\alpha}$, the herd coefficient $\eta>\breve{\eta}$, whereas if $\alpha>\breve{\alpha}$, then $\eta<\breve{\eta}$. The regulator's optimal policies and compensations, the follower's optimal decisions at the terminal time $T$, and the economic gains with and without the regulatory cost are in Fig. \ref{fig:case1}. For comparison, we plot the follower's optimal decision in the unregulated scenario. From Fig. \ref{fig:case1}, the follower's risk coefficient can be divided into three regions: (i) $\alpha\geqslant\tilde{\alpha}$, (ii) $\breve{\alpha}\leqslant\alpha<\tilde{\alpha}$, and (iii) $\alpha<\breve{\alpha}$.

\begin{itemize}
    \item For region (i), the follower is more risk-averse than the leader, and herding increases the social welfare. Hence, the optimal policy is zero, and the optimal policy compensation is $v^{-1}(\chi)$, which is equal to $0$ in this case. Additionally, the follower's decision coincides with the unregulated scenario, and the economic gain is zero, regardless of whether the regulatory cost is considered. 
    \item For region (ii), when considering regulatory cost, it outweighs the economic gain, leading to the same conclusions as in region (i), whereas without regulatory cost, the regulator intervenes with $q^*(\eta)=\eta$, and provides a larger compensation $c^*(\eta)$. 
    \item For region (iii), regardless of whether the regulatory cost is considered, the regulator's optimal policy is $q^*(\eta)=\eta$, and the optimal compensation $c^*(\eta)$ increases as $\alpha$ decreases. Since the follower is more risk-taking than the leader, reducing the herd coefficient through regulation leads the follower to invest more in the risky asset than in the unregulated scenario, and the economic gain increases as $\alpha$ decreases.
\end{itemize}

The above results validate our analysis in Section \ref{sec:analysis}.

\begin{figure}[H]
    \begin{center}
        \includegraphics[width=0.5\linewidth]{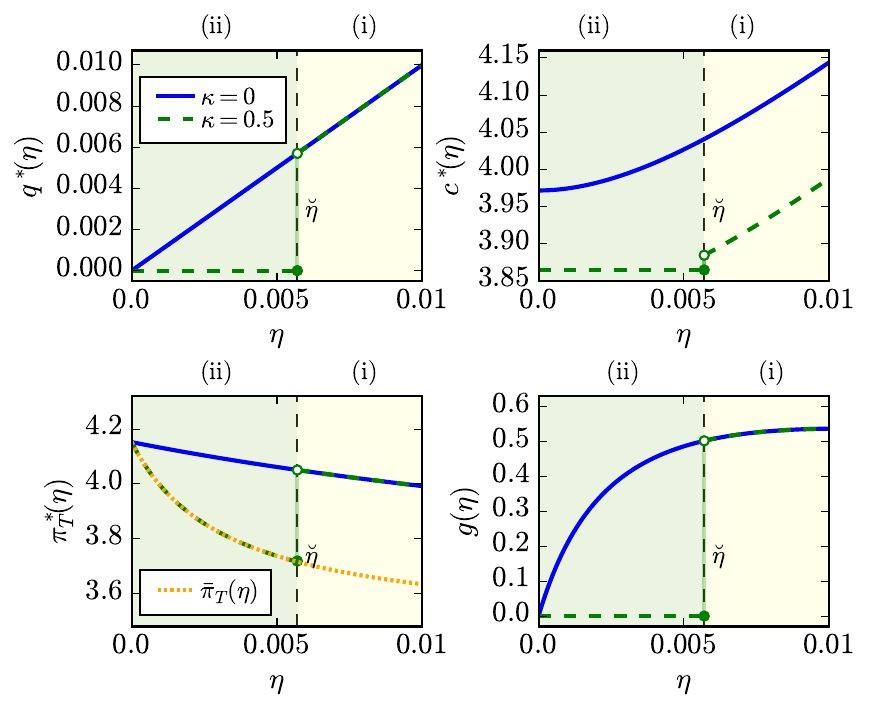}
        \vspace{-6mm}
        \caption{The regulator's optimal policy and compensation, the follower's optimal decision, and the economic gain when the follower's risk coefficient $\alpha=0.25$.} 
        \label{fig:case2}
    \end{center}
\end{figure}

\subsection{Case 2 $({\alpha=0.25})$}
Next, we fix the follower's risk coefficient at $\alpha=0.25<\tilde{\alpha}$ and change the herd coefficient $\eta$. We calculate the herd coefficient threshold $\breve{\eta}\approx0.006$. The experiments for $\alpha\geqslant\tilde{\alpha}$ are omitted as they are identical under the regulated and unregulated scenarios. The results are in Fig. \ref{fig:case2}, and the follower's herd coefficient can be divided into two regions: (i) $\eta>\breve{\eta}$ and (ii) $\eta\leqslant\breve{\eta}$.

\begin{itemize}
    \item For region (i), regardless of whether the regulatory cost is considered, both the regulator's optimal policy $q^*(\eta)$ and compensation $c^*(\eta)$ increase with the herd coefficient $\eta$. The follower's optimal decision without regulation $\bar{\pi}_T(\eta)$ decreases as the herd coefficient increases, reflecting the negative effect of herding in reducing funds allocated to the risky asset. In contrast, the follower's optimal decision with regulation $\pi^*_T(\eta)$ remains unchanged, since the regulator's policy offsets the rising herd coefficient. Furthermore, the economic gain $g(\eta)$ increases with the herd coefficient.
    \item For region (ii), the conclusion depends on whether the regulatory cost is considered. With the regulatory cost, the optimal policy is zero, the optimal policy compensation is $v^{-1}(\chi)$, which is equal to $0$ in this case. Additionally, the follower's decision coincides with the unregulated scenario, and the economic gain is zero. Without the regulatory cost, the conclusions align with region (i). 
\end{itemize}

The above results validate our analysis in Section \ref{sec:analysis}.

\end{document}